\documentclass[9pt,twocolumn]{extarticle}
\pdfoutput=1

\usepackage[sort&compress]{natbib}
\setcitestyle{super}
\setcitestyle{square} 
\setcitestyle{comma} 
\setcitestyle{open={}}
\setcitestyle{close={}}

\usepackage{graphicx}
\usepackage{amssymb, amsmath}

\usepackage[T1]{fontenc}

\usepackage{mathpazo} 
\usepackage[margin=1in]{geometry}

\usepackage{caption}
\usepackage{xcolor}
\usepackage{authblk}
\newcommand{\Psplit} {\ensuremath{|\Psi_\text{split}\rangle \,}}
\newcommand{\Pbunch} {\ensuremath{|\Psi_\text{bunch}\rangle}}
\newcommand{\Rate}[2] {\ensuremath{#1 \mkern-4mu\times \mkern-2mu #2}}
\newcommand{\sinc} {\ensuremath{\mathrm{sinc}\mkern-2mu}}

\newcommand{\nm} {\ensuremath{\text{nm}}}

\newcommand{\chithree}{$\chi^{(3)}$\,}

\newcommand{\pseudosection}[1]{\vspace{9pt}\noindent\textbf{#1}}

\newcommand{\ad}[1]{\textsuperscript{#1}\kern-2pt}

\def\mytitle{On-chip quantum interference\\between silicon photon-pair sources}
\title{\vspace{-1.0cm}\Huge\textbf{\textsf{\mytitle}}}

\author{\textsf{J. W. Silverstone\ad{1*}, D. Bonneau\ad{1*}, K. Ohira\ad{2}, N. Suzuki\ad{2}, H.~Yoshida\ad{2}, N.~Iizuka\ad{2}, M.~Ezaki\ad{2}, C.~M.~Natarajan\ad{3}, M.~G.~Tanner\ad{4}, R.~H.~Hadfield\ad{4}, V.~Zwiller\ad{5}, G.~D.~Marshall\ad{1}, J.~G.~Rarity\ad{1}, J. L. O'Brien\ad{1}, M. G. Thompson\ad{1}}}

\date{}

\begin{document}
\twocolumn[{%
\maketitle
\vspace{-5mm}
\begin{center}
\begin{minipage}{0.7\textwidth}
\begin{center}
\textsf{\footnotesize\textsuperscript{1}Centre for Quantum Photonics, H. H. Wills Physics Laboratory \& Department of Electrical and Electronic Engineering, University of Bristol, Merchant Venturers Building, Woodland Road, Bristol BS8 1UB, UK. \\\textsuperscript{2} Corporate Research \& Development Center, Toshiba Corporation, \\1, Komukai Toshiba-cho, Saiwai-ku, Kawasaki 212-8582, Japan. \\\textsuperscript{3} E. L. Ginzton Laboratory, Stanford University, Stanford 94305, USA. \\\textsuperscript{4} School of Engineering, University of Glasgow, Glasgow G12 8QQ, UK. \\\textsuperscript{5} Kavli Institute of Nanoscience, TU Delft, 2628CJ Delft, The Netherlands\\\textsuperscript{*} Authors J.W.S. and D.B. contributed equally to this work.\\}
\end{center}
\end{minipage}
\vspace{+4mm}
\end{center}
}]

\textbf{
Large-scale integrated quantum photonic technologies\cite{PQTOBrien2009, Tanzilli:2011p11516} will require the on-chip integration of identical photon sources with reconfigurable waveguide circuits. Relatively complex quantum circuits have already been demonstrated\cite{ap-sc-qw2010, Shadbolt-np-2011, Metcalf:2013p11449, Tillmann:2013p11479, Crespi:2013p11478}, but few studies acknowledge the pressing need to integrate photon sources and waveguide circuits together on-chip\cite{Nobuyuki-entangledpair-silicon, nice-njp-quantumrelay-2012}. A key step towards such large-scale quantum technologies is the integration of just two individual photon sources within a waveguide circuit, and the demonstration of high-visibility quantum interference between them. Here, we report a silicon-on-insulator device combining two four-wave mixing sources, in an interferometer with a reconfigurable phase shifter. We configure the device to create and manipulate two-colour (non-degenerate) or same-colour (degenerate), path-entangled or path-unentangled photon pairs. We observe up to $\mathbf{100.0\pm 0.4\%}$ visibility quantum interference on-chip, and up to $\mathbf{95\pm 4\%}$ off-chip. Our device removes the need for external photon sources, provides a path to increasing the complexity of quantum photonic circuits, and is a first step towards fully-integrated quantum technologies.
}

\begin{figure}[t]
\begin{center}
\includegraphics[scale=0.9]{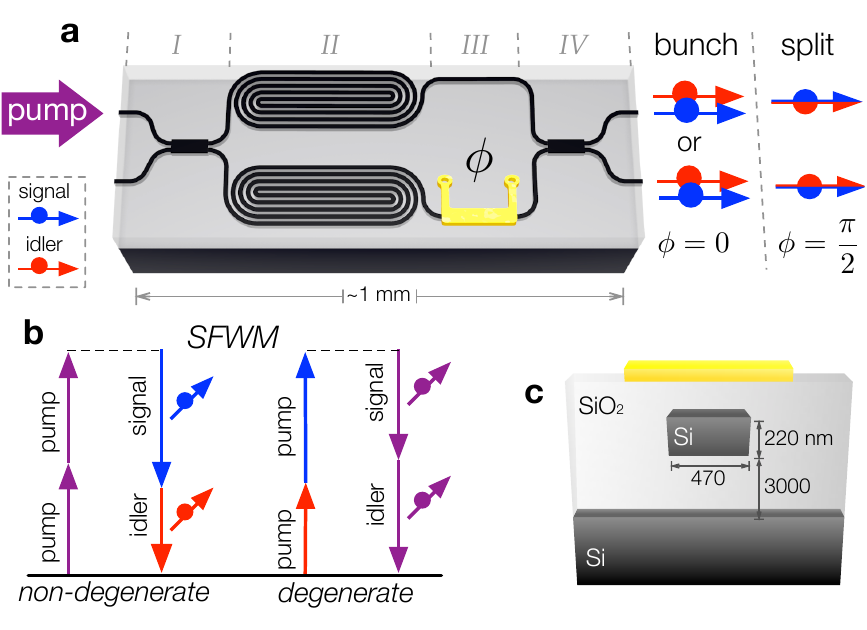}
\caption{\textbf{\normalsize|} \textsf{ Mode of operation, mechanism of photon-pair generation, and physical structure of the device. \textbf a. Schematic of device operation. A bright pump laser is coupled onto the silicon-on-insulator (SOI) chip using a lensed optical fibre and on-chip spot-size converters (not shown). The pump is distributed between two modes via a $2\times 2$ multi-mode interference coupler (\emph{I}), and excites the \chithree spontaneous four-wave mixing (SFWM) effect within each spiraled SOI waveguide source (\emph{II}), to produce signal-idler photon pairs in the two-photon entangled state $\frac{1}{\sqrt{2}} (|20\rangle-|02\rangle)$. The pairs are thermo-optically phase shifted ($\phi$, \emph{III}), and interfered on a second coupler  (\emph{IV}) to  yield either bunching or splitting over the two output modes, depending on $\phi$. \textbf b. Energy diagrams of both types of SFWM, showing the time-reversal symmetry between the non-degenerate and degenerate processes. \textbf c. SOI  waveguide cross-section, with thermal phase shifter on top.}}
\label{operationfig}
\end{center}
\end{figure}

Most quantum waveguide circuits to date have been fabricated from glass-based materials, which offer low propagation loss, a wide transparency window, and efficient coupling to optical fibre, but also limit device functionality and suffer from large circuit footprints. The silicon-on-insulator photonics platform (SOI), recently developed for classical applications\cite{SorefSiRev2006, sun2013}, has several advantages over glass-based systems, including a high component density, a strong \chithree optical nonlinearity, mature fabrication techniques, fast optical modulators\cite{Reed:2010p11508}, and compatibility with both 1550-nm telecom optics and CMOS electronics. As such, SOI quantum photonic circuits \cite{bo-njp-2012} and spontaneous photon-pair sources \cite{ sh-pg-oe-2006, cl-oe-17-16561, Takesue2008Generation, az-oe-ringFWM-2012, Xiong-OL-2011} have recently been shown.

Our SOI photonic device is presented in Fig.~\ref{operationfig}a. Inside the device are two photon-pair sources, each comprising a 5.2-mm-long spiraled silicon waveguide in which the \chithree spontaneous four-wave mixing (SFWM) process is possible (region \emph{II}, Fig.~\ref{operationfig}a). SFWM creates a signal-idler photon pair by annihilating two photons from a bright pump beam (Fig.~\ref{operationfig}b). Non-degenerate pairs are created by a single-wavelength pump, while degenerate pairs require a dual-wavelength scheme\cite{GarayPalmett:2008p11467, PRA-Kumar-2007}. In our experiments, two amplified continuous-wave lasers produced the required pump field. Pump distribution and single-photon interference were achieved using $2\times2$ multi-mode interference couplers (MMI, reflectivity $\approx50\%$), while a thermal phase-shifter modified the on-chip quantum and bright-light states.  Off-chip wavelength-division multiplexers (WDM) were used to separate the signal, idler, and pump channels, before the photon pairs were finally measured by two superconducting single-photon detectors (dark count rate 1~kHz, gate width 650~ps).


Evolution of the degenerate quantum and bright-light states proceeded, referring to regions (\emph{I}-\emph{IV}) of Fig.~\ref{operationfig}a, as follows. The bright pump was equally split by the first MMI (\emph{I}) between the two sources (\emph{II}). By operating in the weak pump regime (so that only one pair was likely to be generated), the simultaneous pumping of both sources yielded a path-entangled $N00N$ state: $\frac 1 {\sqrt 2} (|20\rangle - |02\rangle)$. The relative phase was then dynamically controlled via a thermal phase-shifter in one arm (\emph{III}), which applied a $\phi$ shift to the bright-light pump and a $2\phi$ shift to the entangled biphoton state: $\frac 1 {\sqrt{2}} (|20\rangle - e^{i2\phi}|02\rangle)$. Finally, the bright-light and biphoton states were interfered on a second MMI (\emph{IV}) to yield Mach-Zehnder interference fringes in the bright pump transmission, and half-period ($\frac\lambda2$-like) fringes\cite{Matt-NP-3-346} in the photon pair statistics, of the form
\begin{equation}
|\Psi_\text{out}\rangle =\cos\phi\, \Pbunch +\sin\phi\, \Psplit. \label{eq:outputstate}
\end{equation}
Here, the \Pbunch\ state describes photons bunched together (coalesced) in either output mode $A$ or $B$, and the \Psplit\ state describes pair splitting, with one photon in each mode\cite{PRA-Kumar-2007}. By considering the \emph{degenerate} pair case, and specifically setting $\phi=0$ or $\phi=\frac \pi 2$, we can obtain either state at the output:
\begin{equation}
	|\Psi_\text{out}\rangle = 
\left\{\begin{array}{ll}
	\Pbunch=\frac{1}{\sqrt 2} \big(|20\rangle-|02\rangle\big),& \text{for }\phi = 0\\[5pt]
	\Psplit=|11\rangle, & \text{for }\phi = \frac{\pi}{2}.
\end{array}\right.
\label{eq:outputstatedegen}
\end{equation}

When \emph{non-degenerate} pairs are created, signal-idler exchange symmetry leads to identical quantum evolution and non-classical interference as eqs.~(\ref{eq:outputstate},\ref{eq:outputstatedegen}). The corresponding \Pbunch\ and \Psplit\ states for non-degenerate SFWM are: 
\begin{equation}
\begin{split}
	\Pbunch=\frac{1}{\sqrt{2}}\Big(|1_s1_i\rangle_A |0_s0_i\rangle_B &- |0_s0_i\rangle_A |1_s1_i\rangle_B\Big)\\
	\Psplit=\frac{1}{\sqrt{2}}\Big(|1_s0_i\rangle_A |0_s1_i\rangle_B &+ |0_s1_i\rangle_A |1_s0_i\rangle_B\Big),
\end{split}\label{eq:outputstatenondegen}
\end{equation}
where $s$ and $i$ indicate signal and idler wavelengths (for more detail, see Supplementary Section S4). We will show experimentally that these \emph{non-degenerate} colour-entangled\cite{Ra-dip-pra-1990, DTCE-2009} states behave like \emph{degenerate} pairs.

High-visibility quantum interference---the non-classical interference between two photons\cite{Pittman:1996p11500} on a beamsplitter---underpins all of photonic quantum information science and technology. Quantum interference within our device was quantified by the splitting and bunching probabilities at the output, $P_\text{split}$ and $P_\text{bunch}$, as the on-chip phase ($\phi$) was varied;  classical interference was observed in the transmission of the bright pump through the interferometer. First, we manipulated the on-chip path entanglement of the two-colour (non-degenerate) pairs. The experimental apparatus is pictured in Fig.~\ref{fringesfig}a, and detailed in Supplementary Section S1. WDMs separated the monochromatic 1549.6-nm pump from the non-degenerate signal and idler photons (detuned by $\delta = 6.4\,\nm$) such that signal-idler coincidences and pump transmission could be measured at the same time. Detector coincidences \Rate{A_s}{B_i} and \Rate{B_s}{A_i} were measured for $P_\text{split}$, while \Rate{A_s}{A_i} and \Rate{B_s}{B_i} were measured for $P_\text{bunch}$.  The rates of classical transmission, as well as $P_\text{split}$, and $P_\text{bunch}$ are recorded in Figs.~\ref{fringesfig}b, \ref{fringesfig}c, and \ref{fringesfig}d, respectively. In all cases, high-visibility interference was observed, and both the $P_\text{split}$ and $P_\text{bunch}$ two-photon fringes exhibited a phase-doubling compared to their classical counterparts---a signature of path-entangled two-photon $N00N$ states.

\newpage
\begin{figure}[h!]
\begin{center}
\includegraphics[scale=0.9]{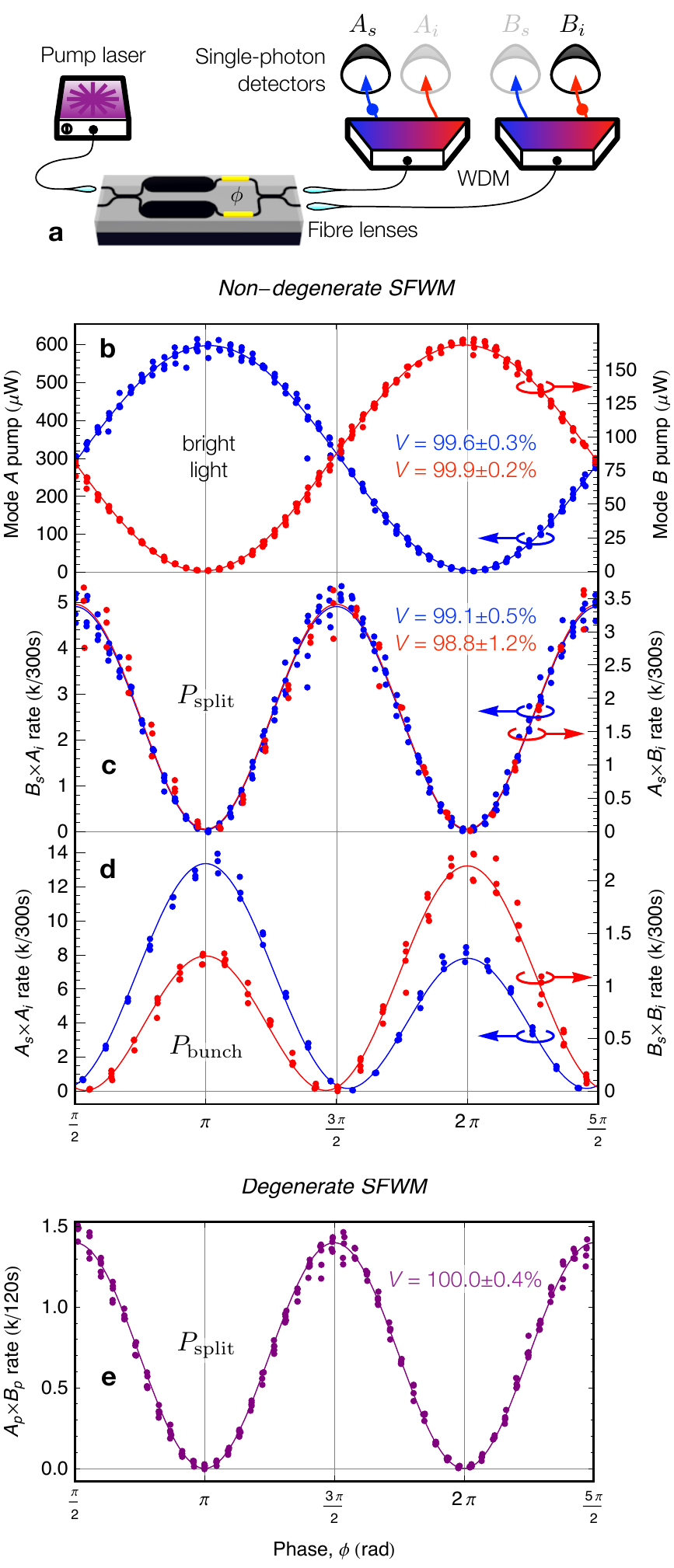}
\caption{\textbf{\normalsize|} \textsf{ On-chip quantum and classical interference measurements, varying the internal phase $\phi$. Coincidence data have accidental coincidences subtracted. \textbf a. Apparatus, showing the coupling of light from a bright pump laser into the device, via fibre lenses, and the separation of signal (blue), idler (red), and pump (violet) wavelength channels using wavelength-division multiplexers (WDM). \textbf b. Transmission of the bright pump laser, showing classical interference. \textbf c. Measurement of signal-idler photon splitting between modes $A$ and $B$, showing quantum interference. \textbf d. Measurement of signal-idler photon bunching, with signal and idler both in mode $A$ or mode $B$. Fringe asymmetry arises from spurious SFWM pairs generated in the input and output waveguides (see text). \textbf e. Photon splitting, like (b), but with monochromatic photon pairs, created via degenerate SFWM.}}\label{fringesfig}
\end{center}
\end{figure}
\newpage

According to equation~\eqref{eq:outputstate}, the splitting rate should follow $P_\text{split}\propto\sin^{2}\phi$ (curves, Fig.~\ref{fringesfig}c). This model works well, with the data exhibiting net visibilities of $99.1\pm0.5\%$ for \Rate{A_i}{B_s}, and $98.8\pm1.2\%$ for \Rate{A_s}{B_i}, where we compute the visibility $V=(N_\text{max}-N_\text{min})/N_\text{max}$ from the maximum $N_\text{max}$ and minimum $N_\text{min}$ values of each fit. The uncorrected visibilities for these measurements remained around 96\%---all raw visibilities, including those for the following measurements, are listed in Supplementary Section~S7. Meanwhile, for the bunching rate, eq.~(\ref{eq:outputstate}) predicts $P_\text{bunch}\propto\cos^{2}\phi$, but an asymmetric behaviour is observed instead (Fig.~\ref{fringesfig}d). This behaviour can be explained by considering spurious SFWM in the device's input and output (I/O) waveguides---identical in cross-section to the source waveguides (Fig.~\ref{operationfig}c). We calculate corrected bunching probabilities, $P_\text{bunch}^A$ and $P_\text{bunch}^B$, at the two output modes $A$ and $B$ as,
\begin{equation}
P^{\substack{A \\ B}}_\text{bunch}=\big|(\Gamma_0+\Gamma_\text{I/O})\cos\phi\mp\Gamma_\text{I/O}\big|^{2}
\label{corrprobeq}
\end{equation}
where $\Gamma_0$ is the base SFWM rate of the spiral sources, and $\Gamma_\text{I/O}$ quantifies the generation rate of spurious pairs inside the I/O waveguides (see Supplementary Section~S5). Equation~\eqref{corrprobeq} describes the $P_\text{bunch}$ data well (curves, Fig.~\ref{fringesfig}c). We extracted the spurious pair rate $\Gamma_\text{I/O}^2$, and found that such pairs accounted for a small fraction of the total: $\Gamma_\text{I/O}^2/\Gamma_{0}^2=2.5\%$ for $P^A_\text{bunch}$, and 2.1\% for $P^B_\text{bunch}$. Since SFWM efficiency scales quadratically with interaction length ($\Gamma^2 \propto L^2$, see Supplementary Section~S3), we compared these ratios with the I/O-to-source waveguide-length ratio squared, $L_\text{I/O}^2/L_0^2 =  2.0\%$, and found good agreement. The $\Gamma_\text{I/O}^2/\Gamma_0^2$  ratio is of considerable importance---it measures the amount of \Pbunch\ contamination when the device is configured to produce only \Psplit. In our experiments, spurious pairs limited the observable off-chip quantum interference visibility to $V<98\%$ (Supplementary Section~S6). This noise could be suppressed by modifying the waveguide geometry outside the source regions\cite{cl-oe-17-16561}, or by moving to resonant sources\cite{az-oe-ringFWM-2012}.

\begin{figure}[h!]
\begin{center}
\includegraphics[scale=0.90]{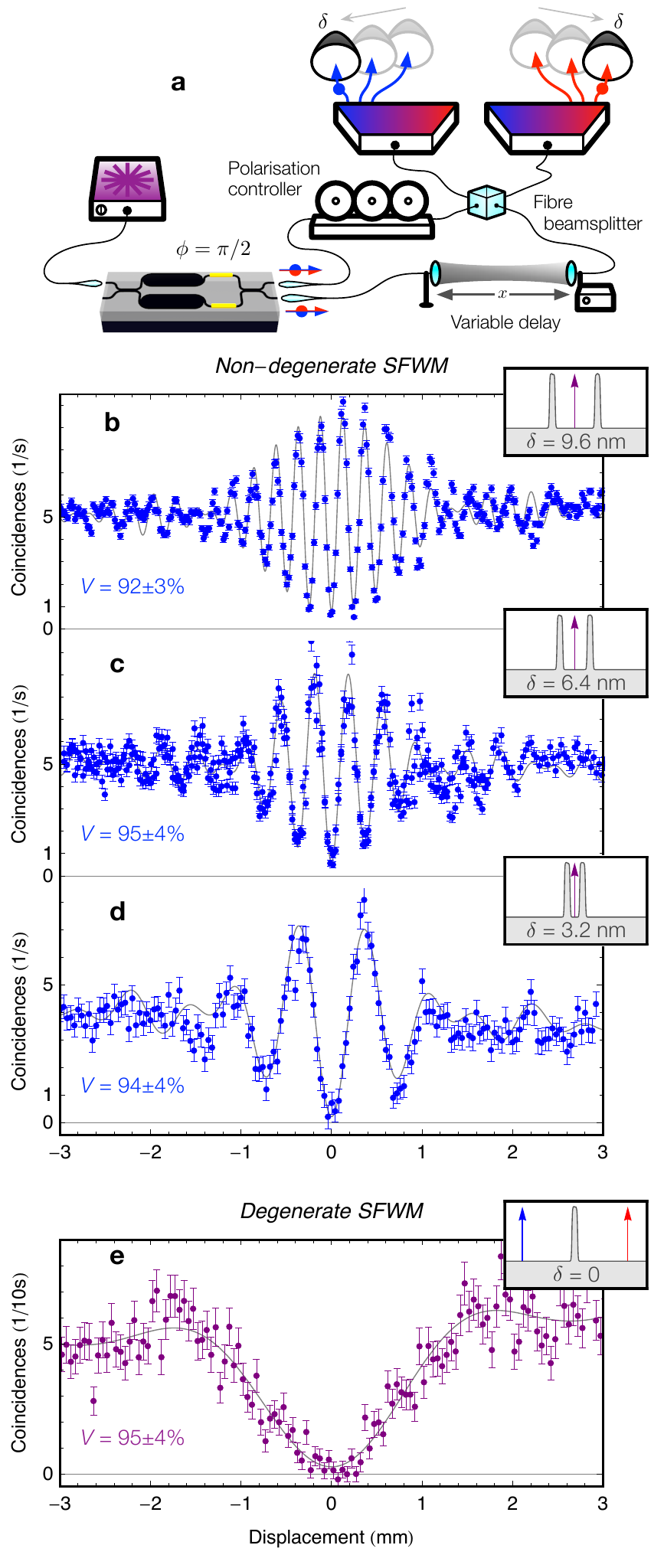}
\caption{\textbf{\normalsize|} \textsf{ Off-chip Hong-Ou-Mandel quantum interference measurements of \Psplit. Beating within each fringe is explained by the signal-idler detuning, $\delta$, as plotted in insets to \textbf b - \textbf e. Coincidence data have accidental coincidences subtracted. Error bars are Poissonian, based on raw coincidences. \textbf a. Experimental schematic. Photon pairs in the \Psplit\ state exit the chip, and one is delayed by a displacement $x$, while the other is polarisation matched, then the pair is interfered on a beamsplitter. Two detectors measure coincidences at different signal-idler detunings $\delta$. \textbf b. Detuning $\delta = 9.6\,\nm$. \textbf c. Detuning $\delta = 6.4\,\nm$. \textbf d. Detuning $\delta = 3.2\,\nm$. \textbf e. Degenerate SFWM, no detuning.}}
\label{dipsfig}
\end{center}
\end{figure}

To show how our \Psplit\ pairs could be used in an external circuit, and to explore the implications of the colour entanglement of our non-degenerate pairs, we performed several off-chip Hong-Ou-Mandel-type (HOM) quantum interference measurements\cite{Hon-PRL-59-2044}, for various values of the signal-idler detuning $\delta$. The modified apparatus is pictured in Fig.~\ref{dipsfig}a. We configured the device to generate \Psplit\ at the chip output (by setting $\phi=\pi/2$), then sent one photon to a tunable delay line, and the other to a polarisation controller. Thus, the optical path and polarisation of the two photons could be precisely matched, and we could introduce distinguishability in the arrival time degree of freedom. The photon pairs then interfered on a fibre beamsplitter ($R=50.2\%$), and we recorded coincidences while varying the arrival time (via the free-space displacement, $x$).

The phase-stable two-colour \Psplit\ state yielded HOM interference fringes (Fig.~\ref{dipsfig}b-d) which exhibited a beating between the non-degenerate signal and idler wavelengths\cite{Ou:1988p11462, Ra-dip-pra-1990} (with detuning $\delta$). Due to the colour entanglement present in \Psplit, the filtered biphoton spectrum has two lobes---one lobe from each of the signal- and idler-channel filters (insets Fig.~\ref{dipsfig}b-d). Since the time-domain HOM interference pattern is effectively the Fourier transform of this spectrum, we can calculate the coincidence probability after the beamsplitter to be

\newpage
\newpage
\begin{equation}
P_\text{HOM}=\frac 1 2 - \frac V 2 \cos\left(2\pi x\frac{\delta}{\lambda_\mathrm{p}^{2}}\right) \sinc\left(2\pi x\frac{w}{\lambda_ \mathrm{p} ^{2}}\right),\label{eq:HOMtwocolor}
\end{equation}
where $x$ is the delay displacement, $\lambda_\text{p}$ is the pump wavelength, $w$ is the WDM channel width, and $\delta$ is the signal-idler channel spacing. As we tuned $\delta\rightarrow 0$, the beat frequency decreased (Figs. \ref{dipsfig}b-d), as predicted by eq. \eqref{eq:HOMtwocolor}. Off-chip HOM interference visibilities were measured for each value of $\delta$: $V_{9.6\,\text{nm}}=94 \pm 4\%$, $V_{6.4\,\text{nm}}=95 \pm 4\%$, and $V_{3.2\,\text{nm}}=92 \pm 3\%$. These results show that our device would perform well as an on-chip two-photon source.

Multi-photon experiments have traditionally used degenerate photon pairs at (near) visible wavelengths in discrete spatial modes to allow arbitrary multi-pair interference\cite{jwp-interfrev-2012-revmodphys}. As a route towards such experiments on a silicon chip at telecom wavelengths, we produced degenerate photons via \emph{degenerate} SFWM---the time-reversed version of the non-degenerate process---which requires a two-colour pump (Fig.~\ref{operationfig}b, see Supplementary Section~S1 for apparatus details). Using degenerate SFWM, with a detuning of $22.4\ \nm$ between the two pump wavelengths, we observed an on-chip quantum interference fringe with $V=100.0\pm0.4$ (Fig.~\ref{fringesfig}e), and measured off-chip HOM interference with $V=95\pm4\%$ (Fig.~\ref{dipsfig}e). The oscillation-free HOM fringe corresponded to eq.~\eqref{eq:HOMtwocolor} with $\delta=0$.

Using both degenerate and non-degenerate SFWM, we observed high-visibility quantum interference, both on- and off-chip (Figs.~\ref{fringesfig} and~\ref{dipsfig}). We observed visibilities on-chip higher than those measured off-chip, and two explanations exist: first, as mentioned, the off-chip visibility was limited by spurious pair generation in the I/O waveguides; and second, the HOM interference measurements required both the on-chip phase, and the off-chip polarisation to be precisely controlled, and drifts in both parameters reduced the visibility. Regardless, the HOM visibilities were still high, and we view the even higher \emph{on-chip} visibilities as showing great promise for future high-fidelity\cite{la-HiFi-2010} source-circuit integrations.

Despite these high visibilities, the measured count-rates were low, due to poor system efficiency (see Methods), and low source brightness. To extract the source brightness, we examined an isolated spiral waveguide: we measured  $2.7\pm0.4\text{ kHz/nm/mW}^2$ for non-degenerate pairs, and a similar $2.5\pm0.6\text{ kHz/nm/mW}^2$ for degenerate pairs, per units bandwidth and launched power. Corresponding coincidence-to-accidental ratios were 290, and 45 (at $100\text{~kHz}$ generation rates; see Supplementary Section~S2). Brightness can be improved by: further engineering our spiral sources, i.e. optimising the spiral length; or moving to resonant\cite{az-oe-ringFWM-2012, Azzini:2013p11510} or slow-light\cite{Xiong-OL-2011} SFWM enhancement. Even with the current brightness, though, we have shown that our device can fill the on-chip role that external crystal- or fibre-based sources have traditionally taken.

Scalability is the ultimate goal for any integrated quantum system. Bulk crystal spontaneous pair sources have been used successfully in the largest optical quantum information experiments to date\cite{jwp-interfrev-2012-revmodphys}, including the production\cite{PanJ-Nature-8photons-2012} and manipulation\cite{ji-Nat_482-489} of 8-photon entanglement. These spontaneous sources can, in principle, be scaled beyond this 8-photon limit by using the techniques of active multiplexing and feed-forward\cite{Zotter:2011p8530, Mower-PRA-multiplexed-2011}. These schemes will require integrated fast switches, single-photon detectors, a high system efficiency, and many identical and bright on-chip photon-pair sources. We have shown that two identical SFWM sources can be integrated monolithically in a waveguide circuit and made to interfere with high visibility. This takes us one step closer to integrated quantum circuits capable of generating and manipulating large photonic states.

\section*{Methods}
\vspace{-0.3cm}
\footnotesize
\pseudosection{System efficiency.} Using the quadratic non-degenerate SFWM power dependence, we can extract the on-chip SFWM rate, together with the associated channel efficiencies, $\eta_s$ and $\eta_i$, where
\begin{equation*}
	\eta_s = \frac{R_{CC}}{R_i} = -24.2\,\text{dB}\text{,\hspace{2mm}and\hspace{2mm}}	\eta_i = \frac{R_{CC}}{R_s} = -25.5\,\text{dB},
\end{equation*}
and $R_{CC}$, $R_i$, and $R_s$ are respectively the coincidence, idler channel, and signal channel rates. These values quantify the total loss experienced by a photon from its generation to detection at detector $s$ or $i$, respectively. Both detector efficiencies were measured at 8\% ($-11.0$\,dB). From a cut-back measurement on the same wafer we estimate the propagation loss at 4.1\,dB/cm, though much lower loss is possible\cite{Gnan:2008p11506}. Using $\eta_s$ and $\eta_i$, and bright-light measurements, we calculate the combined MMI and facet loss as 7.3\,dB. We were unable to separate the MMI losses from those of the facet, though very low-loss MMI designs exist\cite{Sheng:2012p11505}.

\pseudosection{Optical apparatus.} We used two continuous-wave tuneable lasers to generate the required pump field (one or two colours). These lasers were amplified by an erbium-doped fibre amplifier, to produce a total power of 80~mW (in the two-colour case, this was divided between two spectral peaks), then filtered (50\,dB extinction), and the remaining 15~mW was edge-coupled onto the chip using 2.5-$\mu$m-beam-waist lensed fibres and on-chip polymer spot-size converters. Fibre alignment was maintained using piezo-controlled feedback on the bright pump transmission. Filtering was used to both clean the pump before the chip, and to separate the pump from the single photons afterwards. See Supplementary Section~S1 for further details.

\pseudosection{Off-chip HOM fringe optimisation.} The apparatus was configured as in Fig.~\ref{dipsfig}. We used the interferometer, formed between the chip and the fibre beamsplitter, to determine the zero-path-length-difference point on the free-space delay. We input amplified noise to this interferometer, and observed white light fringes as we varied the delay, with fringe maxima at the zero-difference point (and also the HOM fringe centre). We then optimised the polarisation controller using the classical fringe visibility.

\pseudosection{Electrical apparatus.} An electrical probe was used to electrically interface between the chip and an ultra-low noise DC power supply. We applied voltages up to 2.3~V to our thermo-optic phase modulator, for a maximum load of 36~mA. Similar devices exhibited fuse voltages around 2.7~V. We correlated electrical power with thermo-optic phase, with good results (Fig.~\ref{fringesfig}b, Supplementary Figure~3).

\pseudosection{Device fabrication.} The silicon nanowire waveguides were fabricated from a silicon-on-insulator (SOI) wafer having a 220~nm slab thickness. The waveguides were designed to be single-moded, having a width of 470~nm with a silicon dioxide upper cladding. I/O coupling was achieved using spot-size converters comprising a 300-$\mu$m-long inverse-taper with a 200-nm-wide tip beneath a $4\times4\ \mu\text{m}^2$ polyimide waveguide. Structures were defined by deep~UV photolithography (248~nm) and dry etching.


\section*{Acknowledgements}
The authors thank A. Politi for useful discussions, and F. Melloti for experimental assistance. This work was supported by the Engineering and Physical Science Research Council (UK), the European Research Council, the Bristol Centre for Nanoscience and Quantum Information, and the European FP7 project QUANTIP. J.W.S. acknowledges support from the Natural Sciences and Engineering Research Council of Canada. R.H.H. acknowledges a Royal Society University Research Fellowship. V.Z. acknowledges support from the Dutch Foundation for Fundamental Research on Matter. G.D.M. acknowledges the FP7 Marie Curie International Incoming Fellowship scheme. J.L.O'B. acknowledges a Royal Society Wolfson Merit Award. M.G. Thompson acknowledges support from the Toshiba Research Fellowship scheme.

\section*{Author contributions}
Authors J.W.S. and D.B. contributed equally to this work.
J.W.S., D.B., J.G.R., J.L.O'B., and M.G. Thompson conceived of and designed the experiments.
J.W.S., D.B., and M.G. Thompson analyzed the data.
K.O., N.S., H.Y., N.I., and M.E. fabricated the device.
R.H.H., V.Z., C.M.N., and M.G. Tanner built the single-photon detector system.
J.W.S., D.B., and G.D.M. performed the experiments.
J.W.S., D.B., G.D.M., J.G.R., J.L.O'B., and M.G. Thompson wrote the manuscript.

\section*{Competing financial interests}
J.W.S., D.B., J.L.O'B., and M.G. Thompson declare UK patent application number 1302895.6. The authors declare no other competing financial interests.

\end{document}